GreenMail: Reducing Email Service's Carbon Emission with Minimum Cost

THESIS

Presented in Partial Fulfillment of the Requirements for the Degree Master of Science in the Graduate School of The Ohio State University

By

Chen Li, B.S.

Graduate Program in Computer Science and Engineering

The Ohio State University

2013

Master's Examination Committee:

Christopher Charles Stewart, Advisor

Xiaorui Wang

Abstract

Internet services contribute a large fraction of worldwide carbon emission nowadays, in a context of increasing number of companies tending to provide and more and more developers use Internet services. Noticeably, a trend is those service providers are trying to reduce their carbon emissions by utilizing on-site or off-site renewable energy in their datacenters in order to attract more customers. With such efforts have been paid, there are still some users who are aggressively calling for even cleaner Internet services. For example, over 500,000 Facebook users petitioned the social networking site to use renewable energy to power its datacenter [1]. However, it seems impossible for such demand to be satisfied merely from the inside of those production datacenters, considering the transition cost and stability. Outside the existing Internet services, on the other hand, may easily set up a proxy service to attract those renewable-energy-sensitive users, by 1) using carbon neutral or even over-offsetting cloud instances to bridge the end user and traditional Internet services; and 2) estimating and offsetting the carbon emissions from the traditional Internet services. In our paper, we proposed GreenMail, which is a general IMAP proxy caching system that connects email users and traditional email services. GreenMail runs on green web hosts [2] to cache users' emails on green cloud instances. Besides, it offsets the carbon emitted by traditional backend email services. With GreenMail, users could set a carbon emission constraint and use traditional email service without breaking any code modification of user side and email server side.

ii

It is also worth to mention that the basic framework of GreenMail is not specific to email service. Choosing email as our focal point is because routing, storing, and serving email alone causes more than 1M tons of carbon emission each year, even though email is less than 1% of Internet traffic [5]. Besides, generalizing the framework to some general content delivery system is what we expect in the future.

Acknowledgments

The author would like to express his heartfelt gratitude to his supervisor, Prof. Stewart who was abundantly helpful and offered invaluable assistance, support and guidance. Deepest gratitude is also to the other member in the graduate committee, Prof. Wang, for his insightful suggestions.

Vita

2007 to 2011  ............................................... B.S. Computer Science and Technology,

Hangzhou Dianzi University, China

2012 to present ............................................ Graduate Research Associate, Department

of Computer Science and Engineering, The

Ohio State University

Publications

MantisMail: Green Content Delivery for Email. Nan Deng, Chen Li, Christopher Stewart,

and Xiaorui Wang, The Ohio State University. Poster at USENIX Conference on

Networked Systems Design and Implementation. Lombard, IL, 2013

Fields of Study

Major Field:  Computer Science and Engineering

Table of Contents

## List of Tables

List of Figures

Chapter 1:  Introduction

Today more and more applications "talk" to Internet services everyday as well as providing their own services to other applications. However convenient and efficient for developers to use, Internet services do witness a significant growing in terms of their carbon footprint as they become increasingly popular. It has been estimated that by 2020, the annual carbon emission of major datacenters worldwide would likely to exceed the footprint of the entire Netherlands [3]. While individual Internet applications and services like Twitter and Facebook are under public pressure that they should reduce their carbon emission [1,4], public content delivery networks (CDN) providers need to find more value-added features to attract customers, many of which concern the "carbon" impact of services they use [5].

Reducing the carbon footprint of a datacenters has been widely studied in academia: most of efforts are trying to solve the problem inside the datacenters [2，6,7,8], or Internet services [9]. While previous approaches are all concrete work, making changes in either existing datacenter infrastructures or Internet services is not that easy mainly because of two reasons: the transition cost could be too high for the expected benefits to offset, and the new modules dealing with carbon footprint may not be stable or compatible enough to interact with the legacy components, especially during the transition phase. However, some users may simply want to achieve certain carbon emission level that current datacenter and renewable technology cannot suffice.

To address the gap between users' demand to be "green" and current development of datacenters and Internet services, we borrow the idea of CDN, i.e. caching frequently accessed content near to end users, proposing a framework of independent operated Internet service that is running under certain carbon constraints, while maintaining the low cost. Directed by this idea, we design GreenMail as our proof-of-concept prototype system. It is an IMAP proxy running on green web hosts [2] and sitting between email client software (e.g. Thunderbird or IPhone built-in email client) and IMAP servers from traditional email service provider like Gmail and Hotmail. Green hosts are defined as web-hosting services providing cloud instances operated under some carbon constraints. Using green host instances, any operation happened inside the GreenMail is guaranteed to meet certain carbon emission level, e.g. carbon neutral. Users bind their email account to GreenMail, linking the backend email server at traditional email service provider side to our IMAP proxy. They access their emails through GreenMail. GreenMail reduce carbon footprint by caching the recent accessed emails on green host instances; upon another user access on the same email, GreenMail directly dispatches the local copy of the email to the user. If the email is not exist in local cache, then GreenMail retrieves the email from "dirty" remote server, cache it and purchases renewable energy credits (REC) to offset the estimated carbon footprint occurred by retrieval operation, if necessary.

There are two ways by which GreenMail capping its carbon footprint:

(1). Operate on green hosts, so that the carbon emissions caused by local operations (i.e. operations inside GreenMail) is neutralized. By using caching mechanism, most operations are localized.

(2) Purchase renewable energy credits from energy markets, so that the estimated carbon emission from remote email servers is neutralized.

The research challenges here are also pretty straightforward:

(1) We need to answer the following question: At given point of time, how many instances should we keep for our system and how many REC do we need to buy to meet both carbon constraint and cost constraint? We could of course keep as many as possible instances to capture almost all the data in memory to make our response time fair quickly; or buy enough REC to make it unlimitedly "green". But we cannot forget both green instances and REC cost money. To sufficiently answer the above question, an optimization model to characterize the cost is desired.

(2) In order to offset the carbon footprint caused by retrieving content from dirty servers, our system must be able to find a reasonable way to estimate carbon emission of remote servers and network links that it has limited knowledge on.

The contributions of this work are three folds:

(1) Mathematically characterize the cost to operate a green CDN-like caching system.

(2) Build a prototype system GreenMail, analyzing the system's cache behavior against both synthetic and real-world load.

(3) Properly estimate carbon emission occurred during the email transmission phase when cache system retrieves content from dirty servers.

Chapter 2: Related Work

This paper models a type of Internet services which bridges existing Internet services while maintain a low cost. Our model studies how to find a configuration to minimize the cost and be efficient as well. In this section, we outline related work in these areas.

Given the intermittent natural of renewable energy, integrating on-site renewable energy generators to datacenters is challenging. Internet services may reduce their carbon footprint by deciding when, if ever, to process requests. Or a service may drop requests and turn off machines (or put them into low-power state) to use less dirty energy. Blink [8] proposed a key-value storage service that transferred popular keys away from nodes that were turned off during intermittent clean energy outages. The challenge was to serve as read and write requests as possible using only resources powered by renewable energy. Li et al. [11] turned off processor cores to increase the ratio of renewable energy to dirty energy on a system. Similarly, Gmach et al. [12, 13] found that server-power capping and consolidation to power servers under low renewable-energy production could enable renewable powered services, albeit with a performance penalty. Deng et al. [6] studied the impact of different approaches to integrate grid ties to power delivery systems inside datacenters. All works mentioned above require more or less changes inside datacenters' infrastructures.

To control the carbon footprint, a service provider could also distribute user requests to datacenters with higher renewable energy production among multiple datacenters that are geographically located in different places. Le et al. [14] studied services that capped their carbon footprints either by cap-and-trade, cap-and-pay or absolutely capped policies. Their key insight was that a central load balancer could route requests between green and dirty datacenters worldwide to maintain a low carbon footprint while meeting SLAs. Zhang et al. [10] studies services that tried to minimize the carbon footprint of certain requests within a fixed budget. Liu et al. [15, 16] proposed a model to assess a datacenter's performance in terms of carbon footprint efficiency. They use weighted linear models to find the best host, proposing a scalable algorithm to do so. The above works all requires changes either within existing Internet services or datacenters.

Our work differs from previous works, as it is an independent Internet service outside existing infrastructures. GreenMail works as an add-on proxy layer to existing email services and users could switch between GreenMail service and traditional email services seamlessly. GreenMail's goal is not only to operate under a pre-set carbon footprint, but also to offset the estimated carbon footprint from traditional Internet services that are used but outside GreenMail itself. Besides controlling the carbon footprint, GreenMail has to cost as little as possible to reduce the operating cost.

Chapter 3: Minimizing Cost under Constraints

Our optimization model is to stay carbon neutral and minimize the cost while maintain a relatively low response time. In this section, we first show the optimization formula and then discuss the possible optimal solution under difference circumstances.

3.1. Optimization Model

The goal of our optimization model is to minimize the capital cost. The cost of GreenMail consists of two parts: 1) The cost to buy renewable energy credit (REC), and 2) the cost to buy green instances from green hosts. Note that the instances are usually paid on demand. For example, in GreenQloud, a green cloud provider, a medium Linux instance with 1 core, 1GB RAM and 40GB disk storage costs $0.036/hour, or $26.28/month [17].

Before purchasing any renewable energy credits, we need to estimate the possible carbon emission needs to be offset. A user request is first delivered to GreenMail through normal TCP connection, which may cause carbon emissions along with this route. If the requested email is in the GreenMail's cache, it is considered as a cache hit, then the email is sent back to the user and there is no other carbon emissions, except the emission along the link back to the user; otherwise, it is a miss, and then the GreenMail retrieves the email from remote email server, which causes carbon emission from the remote email server and the round-trip network link. Figure 1 shows an overview of GreenMail.

Suppose the user requests arrival rate is λ, given a time period T, the carbon emissions consists of the following parts (the unit is normalized to Joules):

- The emission of the link between the user and GreenMail: λTu, where u is the carbon emission along the link of one request.

- The emission of the remote server and the link between remote server and GreenMail: λT(G+ H)M(N), where G is the carbon emission along the link of one request; H is the carbon emission from remote server of one request; M(N) is the miss rate of the GreenMail cache given N instances from green host.

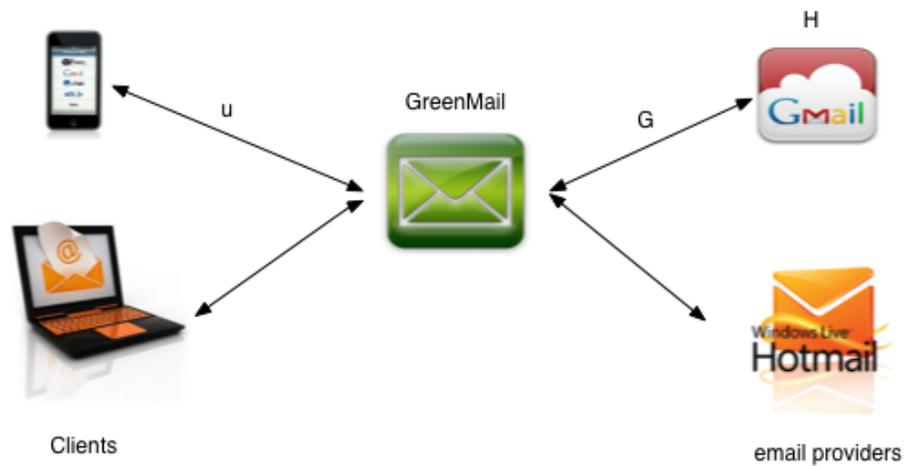

Figure 1 GreenMail Overview

Beside these emissions, there may be additional renewable energy from green host instances to offset some part of the emissions described above. GreenMail works on green hosts' instances. For most green hosts, they lease their instances under some prefixed offset ratio [2]. Here the offset ratio means the ratio of total renewable energy (or renewable energy credits) to the total energy consumption. For example, a 100% offset ratio means the instances only consume renewable energy. Because of renewable energy credits, green hosts may buy additional amount of credits from market to over offset their carbon emission, which makes it is possible for the ratio to be greater than 100%. For example, Green Geeks [18] offers 300% offset ratio, meaning for every Joules of energy used, there are two additional Joules of renewable energy, which could be used to offset the carbon emissions from other places.

To utilize the "over offsets" from green hosts, GreenMail pre-set an offset target ratio for the whole system. The part of such "over-offset" will then be distributed to offset emissions. The additional carbon offset is: $NE_v(r - r_T)$, where $E_v$ is the energy consumption of one instance in GreenMail; r is the offset ratio offered by green host; $r_T$ is the operating offset ratio of GreenMail.

Assuming the unit cost of renewable energy credits is c0 ($/J), and then the cost of renewable energy credits is:

$$c_0 \left[ \lambda Tu + \lambda T(G + H)M(N) - NE_v(r - r_T) \right] \quad (1)$$

Besides the cost of renewable energy credits, another part of the cost is the cost on purchasing instances from green hosts. Assuming the unit price of one in- stance is cv, and then given N instances and within time T, the cost of the instances is: $Nc_vT$ (2)

The total cost of GreenMail is the summation of these two parts of cost is:

$$c_0 \left[ \lambda Tu + \lambda T(G + H)M(N) - NE_v(r - r_T) \right] + N c_v T \quad (3)$$

To ensure service quality of certain SLA, GreenMail has to maintain its processing rate above certain level. Suppose for each instance, the processing rate is β (req/s), and then we have the following constraint for GreenMail is:

$$\beta N >= \lambda \quad (4)$$

meaning the processing rate of the whole system (βN) should be greater than or equal to the arrival rate (λ).

In conclusion, the optimization model for GreenMail is shown in (5). The notations used in (5) are listed in Table 1. Note the notations followed by a "*" means we need to estimate reasonably, which we will discuss in coming sections.

Minimize: $c_0[\lambda Tu + \lambda T(G + H)M(N) - NE_v(r - r_T)] + Nc_vT$, Subject to: $\beta N \geq \lambda$ (5)

| Notations | Descriptions |
|---|---|
| $N$ | Number of instances |
| $T$ | Time period |
| $\lambda$ | Arrival rate of user requests |
| $\beta$ | Processing rate of our system |
| $u *$ | Carbon emission per request on the link between client and GreenMail |
| $G *$ | Carbon emission per request on the link between GreenMail and email servers |
| $H *$ | Carbon emissions per request for remote email servers |
| $M(N)$ | Miss rate function that characterize the GreenMail cache |
| $c_0$ | Unit cost of REC |
| $c_v$ | Unit cost of instances |
| $E_v$ | Energy consumption of one instance in GreenMail |
| $r$ | Offset ratio offered by green host |
| $r_T$ | Operating offset ratio of GreenMail |

Table 1 Notations of Optimization Model

## 3.2. Brief Analysis on the Optimization Model

The optimization model shown in (5) is a non-linear optimization model because of the unknown function $M(N)$. Due to the vicissitudes of the network environment, users' behavior and different cache mechanism, the miss rate function may vary wildly. To solve our optimization problem, we need to at least make some assumptions on the miss rate function. These assumptions will be validated by experiments in Section 5.

Without loss of generality, we can safely make the following assumptions about the miss rate function $M(N)$.

- $M(N) \geq 0$ and $\text{Lim}_{N \to \infty} M(N) = 0$. This is saying the miss rate will never be negative and it reaches zero if we have infinite number of instances.

- $M(N)$ is a monotonically decreasing function. This means with more instances, the miss rate will never be higher.

- $M(N)$ decreases dramatically when $N$ is small; while decreases insignificantly when $N$ is large. This means adding more instances may contribute little to decrease miss rate if we are already using large number of instances. It also suggests that if we have infinite number of instances, adding more instances will not change the miss rate.

All assumptions are made based on our understanding of a typical cache and validated by experiments in Section 5. Some of them may be not always held. For example, under some scenario with certain cache mechanism, the miss rate may increase even if the resources are increased. However, those assumptions will be held in most cases for a well-designed cache given a normal user access pattern.

In this section we first prove the existence of the optimal solution. We consider (3) as carbon constraint and (4) as SLA constraints. We will drop the SLA constraint and prove that the cost function has a minimum point. Then we add the constraint back to get the optimal solution under different conditions.

**Proof that we could always find a global minimum for the cost function:**

To get the global minimum, we calculate the first derivative of (3):

$$\frac{\partial(c_0[\lambda T u + \lambda T(G+H)M(N) - NE_v(r-r_T)] + Nc_vT)}{\partial N} = \lambda T(G+H)c_0\frac{\partial M(N)}{\partial N} + c_vT - E_vc_0(r-r_T) \quad (6)$$

$\lambda, c_0, T, G, H, E_v, r, r_T$ and $c_v$ are all positive values and $r > r_T$. We can then simplify the notation in (3) to:

$$\frac{\partial M(N)}{\partial N} + C_2 \quad (7)$$

Where $C_1 = \lambda T(G+H)c_0 > 0$ and $C_2 = c_vT - E_vc_0(r-r_T)$. If $C_2 \leq 0$, then (6) will always be less than 0, the only constraint we need to satisfy is the SLA, which is $N \geq \frac{\lambda}{\beta}$ according to (4); or when $C_2 > 0$, we could say a minimal point may exist at when $\frac{\partial M(N)}{\partial N} = -\frac{C_2}{C_1}$. However, to prove it is the global minimum, we first need to prove it is a minimum for (3), and then prove there is no other point making the cost function smaller.

We take the second derivative of (3), which is

$$\frac{\partial(C_1\frac{\partial M(N)}{\partial N}+C_2)}{\partial N}=C_1\frac{\partial^2 M(N)}{\partial^2 N}\ (8).$$

Based on the second and third assumptions on the characteristics of *M(N)*, we know $\frac{\partial M(N)}{\partial N}<\frac{\partial M(N_0)}{\partial N_0}$ and $\frac{\partial^2 M(N)}{\partial^2 N}>0$ when $N_0<N$. Thus (8) would always be greater than zero (9). As a result of (9) we could know there would be one and the only one minimal point when N satisfies $\frac{\partial M(N)}{\partial N}=-\frac{C_2}{C_1}$, where $C_1=\lambda T(G+H)c_0>0$ and $C_2=Nc_v-E_vc_0(r-r_T)$.

Finally, consider the SLA constraint. Suppose the $N_{min}$ is the minimum point we get from above:

- If $N_{min}\geq\frac{\lambda}{\beta}$, then $N_{min}$ is the solution to the optimization problem.

- If $N_{min}<\frac{\lambda}{\beta}$, then $\frac{\lambda}{\beta}$ is the solution to the optimization problem.

As a conclusion, we could safely say 1) if $C_2\leq0$, the cost function will always increase with N being greater; and thus the only constraint for N comes from to satisfy the SLA, which is $\frac{\lambda}{\beta}$. 2) If $C_2>0$, we first calculate the minimal N to satisfy the carbon constraint and then check if it could also satisfy SLA constraint: N must at least meet the SLA constraint.

Chapter 4: System Design and Implementation

We leverage several open source software to implement our prototype system. The system includes the GreenMail itself and the synthetic user as well as an IMAP server. Specifically, in GreenMail we use SquirrelMail [19] as front-end from where users could access our service, Redis [20] as our in-memory key-value store used to caching users email. We use Apache Jmeter [21] to simulate the concurrent user requests to the GreenMail web interface and Mstone [22] to conduct the pressure test against MantisMail [5], a pure IMAP proxy version of GreenMail developed by us in another related project. Dovecot [23], an open source IMAP server is used in our system. Figure 2 shows how the above components interact. We will discuss about each component of our system in the following sections.

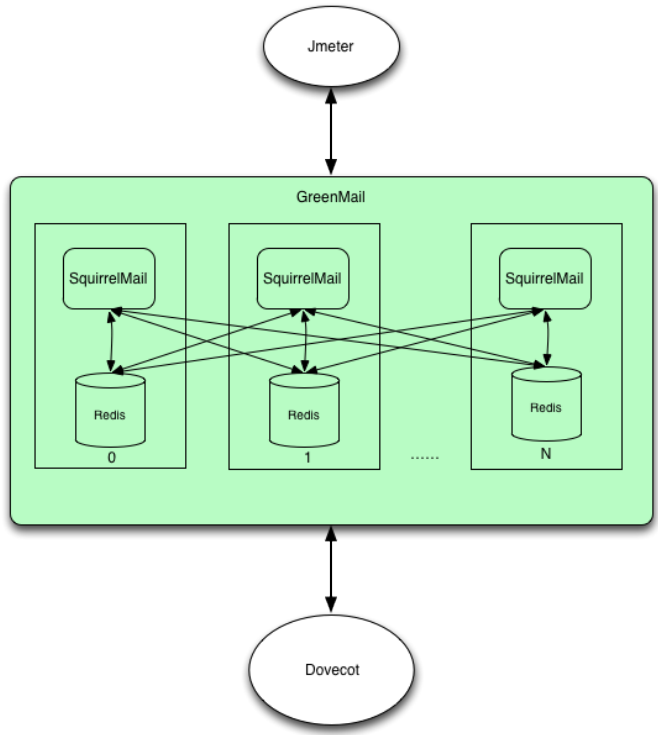

Figure 2 System Design of GreenMail

## 4.1 GreenMail

GreenMail itself consists two parts: front-end and back-end. The front-end is where user interacts with our service and the back-end is where we cache the email.

We implement our front-end by modifying SquirrelMail, an open source webmail package written in PHP, which supports IMAP and SMTP protocol. FETCH commands, which are to retrieve emails content from server in IMAP context, are cached by our system. We let SquirrelMail always first try to retrieve email from Redis; if it is not there then we have to issue request to the IMAP server. Figure 3 shows the pseudo code of this procedure.

```
// First, try to retrieve email
// from Redis
$readback = $redis->get($key);

// Test if the email is in the cache.
if($readback) {
    // If it exists, then return results
    return $readback;
} else {
    //Otherwise, get the email from remote IMAP server
    //and cache the result before returning email
    $read=get_mail_from_IMAP;
    $redis->set($key,$read);
    return $read;
}
```

Figure 3 Pseudo Code of Cache Look-up

The back-end is Redis instances. Redis is an in-memory key-value store for small chunks of arbitrary data (strings, objects) from results of database calls, API calls, or page rendering. By doing caching everything in memory, Redis has significantly reduced the access latency and the load on the persistent storage, which is IMAP server in our case. But in our context, it will also reduce the carbon footprint IMAP server and the link cost to the server.

Redis is typical client-server architecture. Its special designed hash mechanism makes it is pretty scalable, since which instance is chosen to store the item is calculated merely by client side library. It is users' responsibility to ensure each client has the exactly same server list. Redis servers know nothing about each other and no communication is required among server nodes. Redis takes advantage of consistent hashing [24], it minimizes the remapping when the overall hash table resizes. Thus, there is very low miss rate incurred when new nodes are added in or deleted from the cluster.

## 4.2 Other Components

In order to simulate multiple users accessing the email service based on some pattern, we use Apache Jmeter to generate concurrent user requests. We create scripts to randomly generate requests against uniform and Gaussian distributions. The inter-arrival times are Poisson distribution. To conduct pressure test on the system, we also leverage Mstone, a multi-protocol stress and performance measurement tool. We will discuss this in more detail in next section.

To build a controllable environment, we use Dovecot to set up an IMAP server treating it as a remote public email service provider. Since IMAP service is generally available among most email providers like Gmail, Yahoo, GreenMail can seamlessly transit to those major providers.

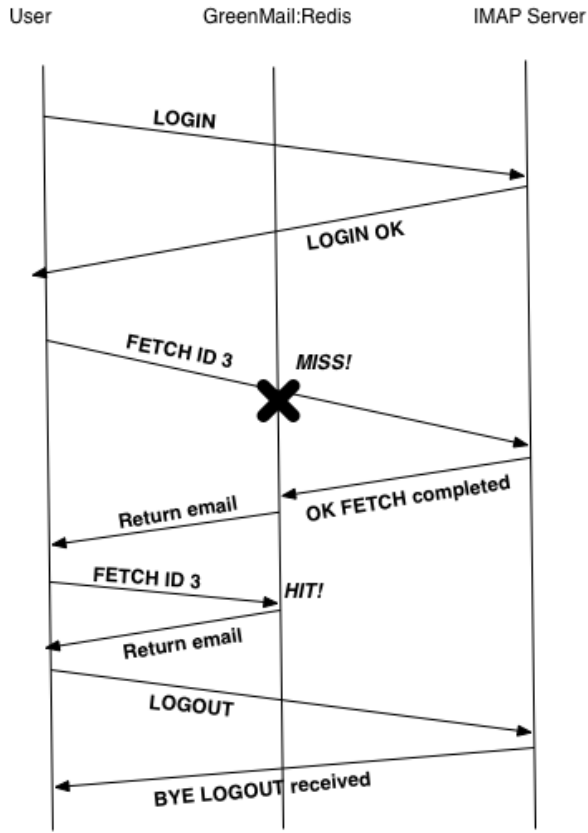

Figure 4 Time Sequences of Typical Accesses to GreenMail

A typical access is shown as in Figure 4. The user first sends an IMAP command of LOGIN to log into the IMAP server, and then opens some email by send FETCH command. Upon intercepting a FETCH command, GreenMail first tries to find the email among Redis clusters. If the corresponding email could be found, then it is a cache hit and there is no need to further resort to IMAP server; Otherwise GreenMail let the FETCH command go through itself. The IMAP server would respond to this FETCH with OK FETCH completed. After caching this email, GreenMail forwards it to the user. Note that since GreenMail is on green host, we will eliminate the possible carbon cost introduced by accessing IMAP server.

As mentioned before, we also need to estimate the carbon footprint outside GreenMail itself, in order to offset the carbon footprint caused by retrieving content from dirty servers. Our model estimates two kinds of carbon emission outside of GreenMail: link cost and server cost.

For server emissions (H in table 1), we use data from Google [25] to get a rough understanding on how much carbon footprint of an email server in different scale (large, medium, small) could be. For example, to serve ~500 people (medium scale), annual energy per user would be 28.4 kWh; and the annul carbon emission per user would be 16.7 kg.

For link cost (G and u in table 1), we need to consider both users to GreenMail (u) and GreenMail to remote servers (G). While u is quite hard to capture since user accesses could happen anywhere from any device; G is relatively fixed given that we only have limited number of IMAP servers and their "IP:Port" are also fixed. However, we could estimate u+G by multiplying average Internet Energy Intensity (kWh/GB) [26] with the inbound traffic (GB). Obviously, such estimation is rough because here we assume the Internet is "flat" in terms of energy consumption. But we could have better estimation if we could have more information on carbon footprint and load of passed routers. For the carbon emission from GreenMail to dirty servers, our system traces each routers from GreenMail to IMAP server, converting IPs to a serial of geo-locations and then according to some published data source, we could know how much of the carbon emission per unit energy at that locations. For example, we could know a router/switch with IP 164.107.2.150 is located at Columbus Ohio, where 1030 Kilograms of $CO_2$ emitted per (net) megawatt-hour of electricity produced. The remaining is just to find some average energy consumption of those routers. For example, from [26], we know that a typical router cost 2.4 TWh per year. While this estimation is pretty rough, our argument is it is enough for us to understand the carbon footprint of the GreenMail and its potential users.

While above estimations make sense in some reasonable way, further mathematical estimation suggests the link carbon cost might be dropped in practice. According to [26], we know in 2006 typical server annual energy consumption is 24.5 TWh and Internet Energy Intensity is 24.kWh/GB. Assume each email user has 50 emails daily, and each of email size is 0.1 MB. We also assume an aforementioned server could serve ~500 users. Then we know the transmission cost would be 22173kWh, which is only 1/1,000,000 of 24.5TWh. In this sense, we could argue that the link cost in our formula could be ignored in our real implementation.

Chapter 5: Experiments

In this section, we will present three sets of experiments on our system: one is using synthetic workload, one is real world test, and the other is a performance comparison between with our IMAP Proxy and without our proxy.

## 5.1 Simulation On Our System

For the synthetic workloads, we use Jmeter to generate user accesses against certain distributions; also, we use Mstone to test how well our system could be when access pressure is high.

### 5.1.1 Characterizing Caching Behavior

In Section 3, we mentioned the three assumptions about the miss rate function $M(N)$. In this section, we set up several experiments to validate those assumptions.

We did our experiment on one Redis instance and configure it to be different sizes, i.e. 4MB, 8MB and 16MB. We do our experiment only on one Redis instance is simply because the goal is to characterize the property of caching behavior with a growing capacity of cache. The tiny capacity of cache in our setup is corresponding to the small size of email content we generated. Email size is against Gaussian distribution with mean 40 KB. There are 10,000 emails in total. Note that, only the 8 MB cache and the 16 MB cache can hold all the accessed items in our test. The user access pattern follows the typical long-tail distribution, suggesting that the majority of requests target small portion of very popular objects. This agrees with the observation that users only re-open most recent emails.

Figure 5 shows the traffic, both incoming and outgoing, of the Redis instances with cache size varying from 4 MB to 16 MB. There are always more traffic from gets than sets since a miss get will cause a set to the cache. And it is also because the Redis client side library compresses the data it sends out to server.

For caches that can hold all the email items, the miss rate will dramatically drop down as time going on and more and more popular items have already been put into cache. The above is not true for the 4 MB cache, which cannot hold all the email in it. As a result, there will always remain certain rate of misses since the old emails are continuously evicting from caching when newer emails are accessed. The same trend is illustrated in Figure 6(a) to Figure 6(c), where the misses/hits per second are shown. For 4 MB cache, we can see it always contain several portion of misses while for other two caches, only tiny portion of requests are missed as time goes on.

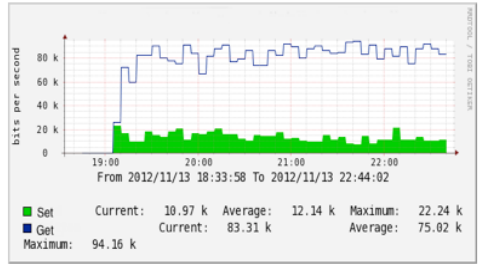

(a) cache size = 4MB

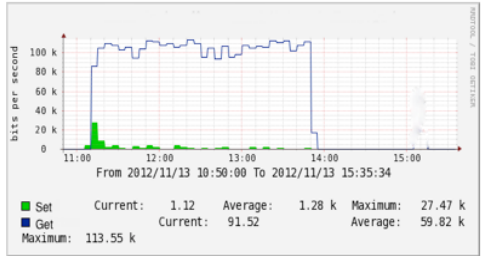

(b) cache size = 8MB

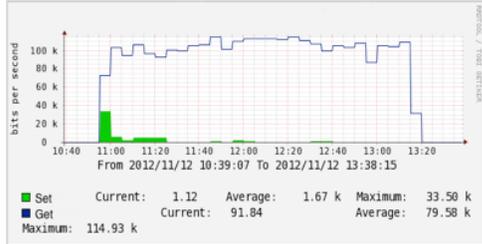

(c) cache size=16MB

Figure 5 Network Traffics

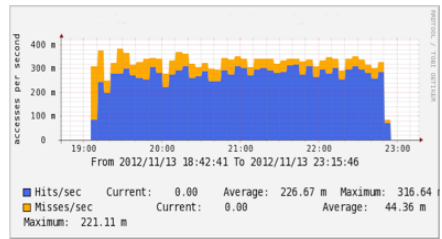

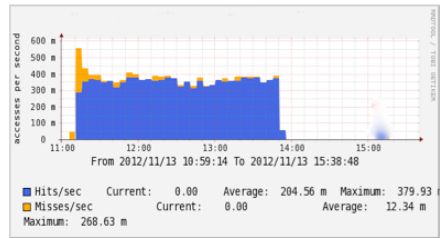

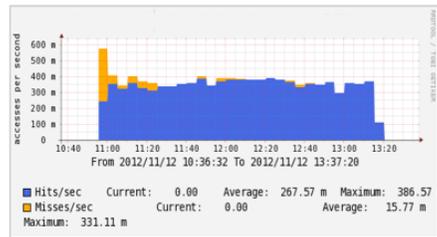

Figure 6 Cache Hits and Misses

In Section 3, we made three assumptions against miss rate function M(N). The above graphs also agree on the assumptions.

First, for assumption 1 and 3, comparing Figure 6(b) to Figure 6(c), when cache size is large enough, we will not see obvious miss rate reduction, especially when compared to Figure 6(a) to Figure 6(b). Intuitively, given fixed amount of user as well as their fixed requests rates, the storage space needed is also fixed. This makes there will be few to no benefits in terms of miss rate decrease since we have already cached everything that may be accessed. Therefore we could safely expect zero miss when cache size is enough large.

Second, *M(N)* is a monotonically decreasing function. Note when cache size grows, fewer cache misses are observed. For 8MB cache and 16 MB cache instance, we barely see a miss after certain time.

*5.1.2 Performance Test against MantisMail*

To get a better understanding on performance difference between with and without our proxy system, we leverage Mstone to conduct performance test against our system. Mstone [22] is a multi-protocol stress and performance measurement tool, which could test multiple protocols including IMAP simultaneously and measures the performance of every transaction.

Figure 7 shows the IMAP commands per second of both w/ vs. w/o our proxy. It suggests that our proxy bring down the system average performance by 42%. This large performance deduction is mainly due to our proxy's current debug version implementation will encrypt, compress and persist the content of each command we cached.

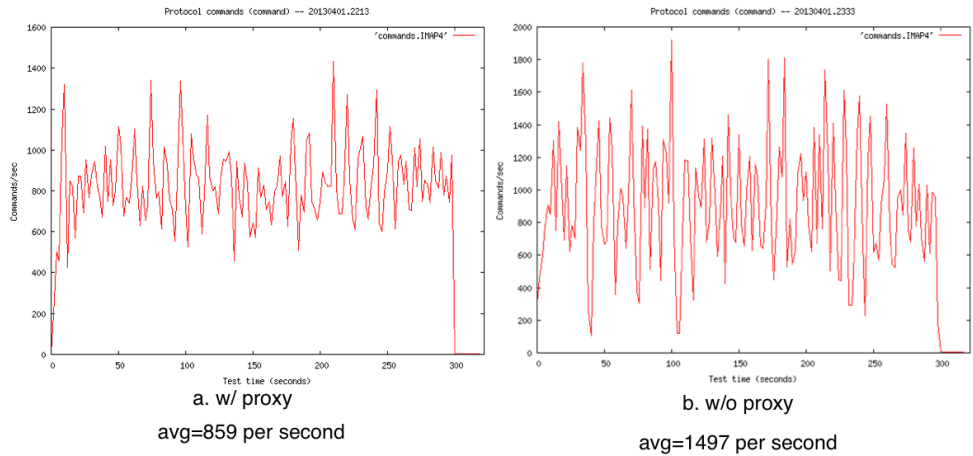

a. w/ proxy
avg=859 per second

b. w/o proxy
avg=1497 per second

Figure 7 IMAP Commands per Second

Table 2 shows the comparison in terms of processing time for each IMAP command. Note that not all types of IMAP command are equally "quick" to process, e.g. LOGIN is slower than most of other IMAP commands. We see the same performance degrade in terms of processing time. With our proxy, its average processing time for w/ proxy is much longer than w/o proxy. Note with proxy, we actually "smooth" the processing rate for each IMAP commands, since we do caching in our proxy. The duration of time to retrieve content from local cache for different commands are similar. That's why we have a smaller standard deviation for w/ proxy.

|  | avg | min | max | std div |
|---|---|---|---|---|
| w/ proxy | 114.54ms | 245.00us | 4.901s | 128.74ms |
| w/o proxy | 65.72ms | 8.00us | 14.995s | 541.35ms |

Table 2 IMAP Command Processing Time

Figure 8 compares throughputs. "Bytes read" means how many bytes are in the data flow from tested system to user; "bytes write" is how many bytes are written out through user (Mstone here). We use server here to denote IMAP server or our caching system; Mstone is client in this context. Figure 8 suggests we always "get" more from server than set to the server. This means we are benefiting from caching frequent accessed objects. Performance degrade appears this figure again for the same reason as mentioned before.

As a conclusion for this experiment, we can say we get benefit by using our system; but the tradeoff is a performance drop. However, we make an argument this degraded performance is mainly because of we did the entire test against our "debugging" version system, which enable a large amount of logging and encrypt and compression.

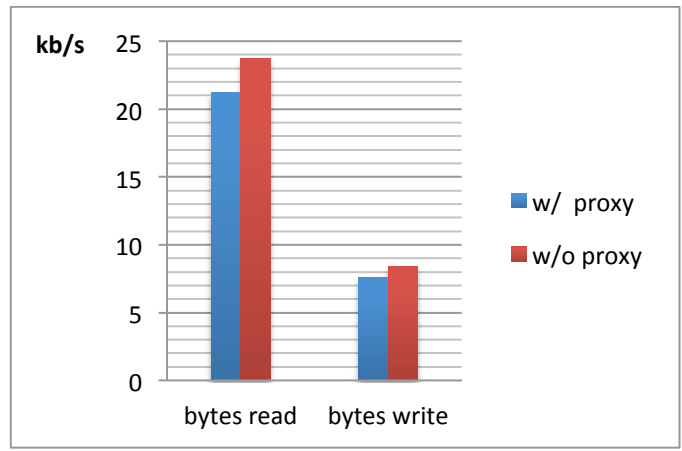

Figure 8 Throughputs

## 5.2 Jump Into Real world

Along with a web interface for user to access their emails, we also have a pure IMAP proxy system that is built under same idea with GreenMail. We argue that actually nowadays people tend to access their email from some mobile devices other than through a web interface. Thus being a part of our experiment to see how our system would act in the real world, we deploy the system in GreenQloud, one of well-known green hosts. And three members of our group including the author himself used the system for two days. We all bind our email clients, such like Thunderbird or Outlook on PC, or iPhone/android built-in email software to the system. Our system logged all the email access and hit/miss activities. Figure 9 shows the share of hits and misses compared to all accesses; Figure 10 shows bytes transferred for all misses and all hits. The numbers here are not that big is because: 1) only three people using the system; 2) The experiment was conducted during weekend instead of busy workdays.

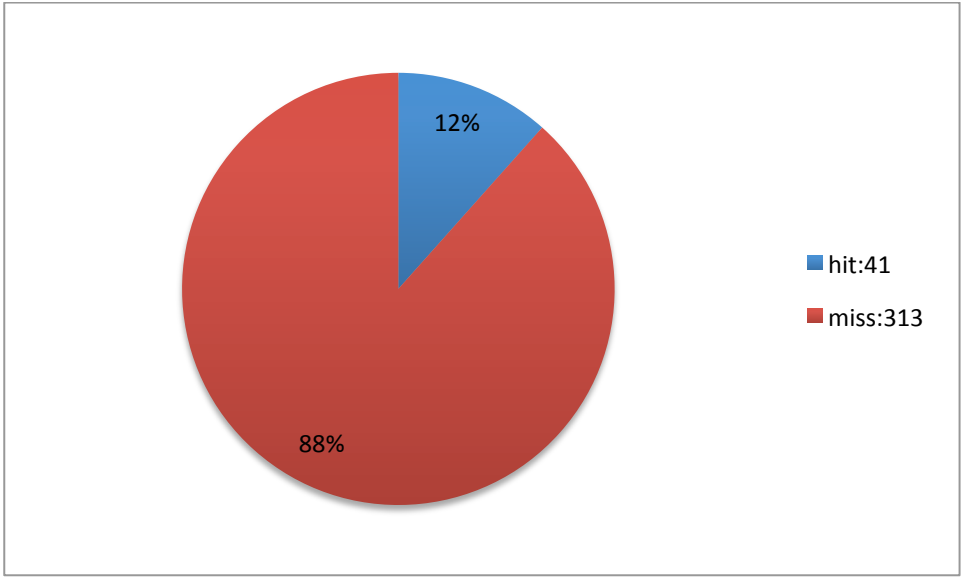

Figure 9 Numbers of Hits vs. Misses

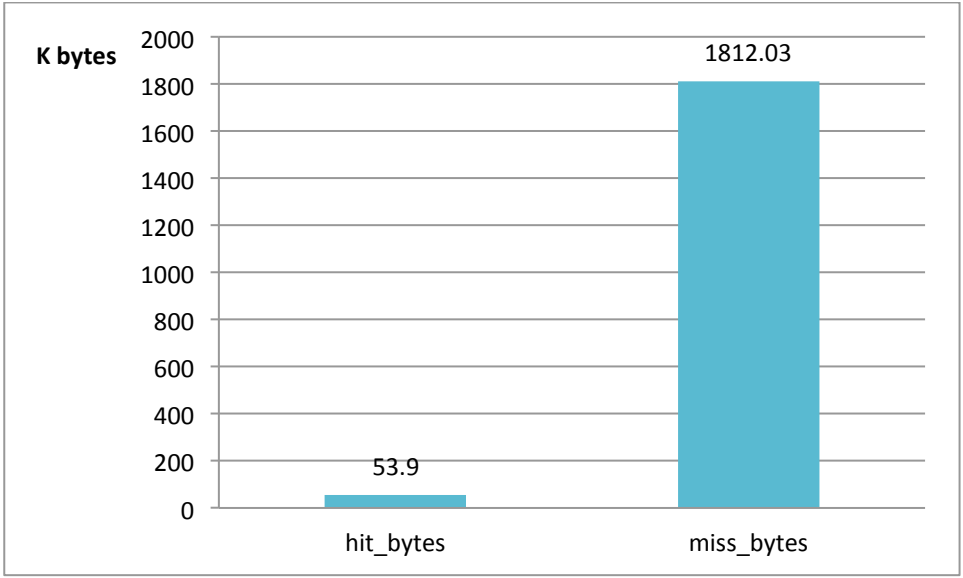

Figure 10 Data Transferred on All Misses vs. All Hits

Figure 9 also shows us that we got 12% of cache hit for all cache accesses in two days, which is actually a pretty big number for a CDN like system. Figure 10 tells the same story in terms of bytes transferred. During the experiment, none of us intentionally or manually open to read an email multiple times. Only the clients automatically connect to pull emails. If we bind more devices under same user, we could expect higher hit rate. It is because most email client software would always periodically check the inbox with different frequency, if one email is accessed by one client and after first miss, the email would be cached and all the following access from other devices would only lead hits. Luckily, what we expect is the way that people accessing their email nowadays. Usually we connect same email accounts from our PCs at work, cellphones, PDAs, laptops, etc.

The fact that ratio of "hit_bytes" to "miss_bytes" is smaller than that of number of hits to number of misses suggests that in terms of size of transferred data we may not further benefit that much from caching other IMAP commands such like LIST, which is used to list the tittle of email in an inbox, even though we may get a higher hit rates if we do so; it is because only the FETCH command get the real content of email, which makes up the majority of network traffic for IMAP communication.

### 5.2.1 Operating Cost Estimation

And then, how much do we need to pay for if we run this experiment system for one year? We suppose our system is carbon neutral, i.e. $r_T$=100%. r is also 100% as GreenQloud is carbon neutral, rather than providing over-offsetted ratio that is greater than 100%. For simplicity, we let N=1. Assume after cache is warmed, the miss rate would be 88% when system is stable as our experiment suggests. Link costs, which include G and u, are ignored. We aggregate the server energy cost by using Google's published data [25], which is 14,200 kWh/year. We let $c_0$ equal to $0.02$/kWh according to [27]. $c_v$ equals to $26.28/month in our experiment[17]. Applying the above to formula (3), the total cost to operate this experiment system is $565.28, where carbon cost is $249.92 and instance cost is $315.36. Note if we could have a lower miss rate, say if users connect multiple devices under same account, the carbon cost would drop.

6. Conclusion

In this paper, we conceptually proposed a framework for CDN-like systems which can attached to existing Internet services, operate under pre-set carbon footprint and offset estimated carbon emissions from the traditional Internet services it used. Based on this idea, we designed and implemented a prototype system called GreenMail, which is an IMAP proxy, reducing carbon emission by caching users' email on green hosts. To get a better understanding on our system, especially the cache system behavior, we thoroughly tested the system against both synthetic and real world workloads, estimating the operating cost of our system for 1 year. We also present how to properly estimate carbon emission occurred outside the GreenMail itself.